\documentclass[12pt,preprint]{emulateapj}

\usepackage{graphicx}
\usepackage{amsmath}

\begin{document}

\title{Hubble Constant from LSST Strong lens time delays with microlensing systematics}
\author{Kai Liao$^{1}$}
\affil{
$^1$ {School of Science, Wuhan University of Technology, Wuhan 430070, China.}}
\email{liaokai@whut.edu.cn}

\begin{abstract}
Strong lens time delays have been widely used in cosmological studies, especially to infer $H_0$.
The upcoming LSST will provide several hundred well measured time delays from the light curves of lensed quasars.
However, due to the inclination of the finite AGN accretion disc and the differential magnification of the coherent temperature fluctuations, the microlensing by the stars can
lead to changes in the actual time delay on the light-crossing time scale of the emission region $\sim days$.
We first study how this would change the uncertainty of $H_0$ in the LSST era, assuming the microlensing time delays
can be well estimated. We adopt 1/3, 1 and 3 days respectively as the typical microlensing time delay uncertainties. The relative uncertainty of $H_0$ will
be enlarged to $0.47\%$, $0.51\%$, $0.76\%$, respectively from the one without microlensing impact $0.45\%$.
Then, due to the lack of understandings on the quasar models and microlensing patterns,
we also test the reliability of the results if one neglects this effect in the analysis. The biases of $H_0$ will be
$0.12\%$, $0.22\%$ and $0.70\%$, respectively, suggesting that 1 day is the cut-off for a robust $H_0$ estimate.

\end{abstract}

\section{Introduction}
Cosmology has entered the precision era with an increasingly huge amount of observation data. The theoretical cosmic models and the parameters therein
have been well studied and constrained. Currently, the most plausible model for describing the Universe is the $\Lambda$CDM model where
the matter content is dominated by cold dark matter and the flat Universe is accelerated by dark energy~\citep{Planck2016}.
Meanwhile, we are facing various tension problems stemming from either the systematics in each measurement or potential new physics.
For example, the Hubble constant ($H_0$) measured by distance ladder method is inconsistent with the one from Planck data by $3\sigma$ uncertainty~\citep{Riess2016,Freedman2017}.
To understand this discrepancy, one way is to measure $H_0$ in a totally independent manner up to sub-percent level with systematics under control~\citep{Weinberg2013}.

Strong gravitational lensing by galaxies~\citep{Treu2010} has been widely applied in studying astrophysics~\citep{Zackrisson2010}
and cosmology~\citep{Treu2016,Liao2017a}. The typical system
consists of a distant quasar lensed by the foreground elliptical galaxy, forming multiple images of the AGN and several pieces of arcs as the imaging of the host galaxy.
The time delays between multiple AGN images measured by comparing their light curves~\citep{Tewes2013}
can be used to constrain $H_0$~\citep{Refsdal1964}, which is not only an independent but also a one-step way comparing with the distance ladder techniques from Type Ia supernovae or the inverse distance ladder techniques from CMB and BAO~\citep{Komatsu2011,Hinshaw2013}. Roughly speaking, the time delay between two AGN images
$\Delta t \propto H_0^{-1}(1-\langle\kappa\rangle)$, where $\langle\kappa\rangle$ is the mean convergence or surface density in the annulus between the images~\citep{Kochanek2002}.
Therefore, the inferred $H_0$ uncertainty should follow the error propagation formula $\sigma_H^2/H_0^2\propto(\sigma_{\Delta t}^2/\Delta t^2+\sigma_{LM}^2+\sigma_{LOS}^2)/N$ for $N$ observed lensing systems. Here we split the convergence uncertainty into
two parts since they are from independent measurements: the lens galaxy modelling uncertainty $\sigma_{LM}$ and the fluctuation along the line of sight (LOS) $\sigma_{LOS}$. The former is determined with the observation of central velocity dispersion of the lens galaxy,
the lensed host galaxy and the AGN image positions~\citep{Treu2002,Suyu2010,Wong2017} while the later comes from the measurements of the line-of-sight mass distribution with spectroscopy and multi-band imaging~\citep{Rusu2017}.

One of the current limitations for the strong lens time delay cosmology is the small number of lensing systems with well-measured time delays, high-resolution imaging and spectroscopy measurements, though a few lenses have shown the power on constraining $H_0$~\citep{Wong2017,Bonvin2017}, (see the state-of-art program H0LiCOW~\citep{Suyu2017}).
A $3\%$ precision of $H_0$ has been
achieved with only 4 lenses~\citep{Birrer2018}. In the next stage, The Time Delay Challenge (TDC)~\citep{Liao2015} shows the upcoming Large Synoptic Survey Telescope (LSST) will bring us 400 well-measured lenses with average precision $\sim 3\%$ and accuracy $\leq1\%$ for the time delay measurements. With these lenses,
we are supposed to achieve an unprecedented precision for $H_0$. However, one must be aware that for any precise measurements, the systematics or bias
should be controlled below the statistical uncertainties to get a robust result, especially for the LSST data which may hit the
systematics floor. Concerns must be addressed for all independent observations, for example, the lens modelling systematics may already dominate over $\sigma_{\langle\kappa\rangle}$~\citep{Schneider2013,Birrer2016}. The ongoing Time Delay Lens Modelling Challenge (TDLMC)~\citep{Ding2018} aims to test the systematics in current algorithms. In this work, we only focus on the time delay measurements and assume the lens modelling and LOS measurements are accurate.

The TDC concluded that the time delay measurements from the light curves are accurate enough compared with the precision. However, recently,~\citep{Tie2018} (hereafter TK18)
found that the time delays measured as the shifts of the light curve pairs in time domain are not the cosmological ones but the combinations of
cosmological and microlensing time delays. The microlensing time delays are due to the differential magnification of the coherent accretion
disc variability of the lensed quasars. This effect can change the measured time delays by light-crossing time scales of the discs $\sim$days.

Therefore, to get an unbiased result,
we need to incorporate this in the analysis. Chen et al. (2018) (hereafter C18) tried to constrain the microlensing effects on time delays and got a weaker constraint for
$H_0$ especially for short time delays. They developed a pipeline in a Bayesian framework using time-delay ratios and simulated microlensing time delay maps as priors.
Birrer et al. (2018) (hereafter B18) estimated the effect of microlensing time delay and found it much smaller than the statistical uncertainty.
However, as pointed, problems still exist due to the assumptions in AGN model, the size of the disc and the local properties of the lens at each image.
One doubts all these assumptions could also bring extra bias unless further blind analysis tests it.
Due to these complexities, Bonvin et al. 2018 chose not to include the microlensing time delays in the estimates if there is no evidence showing the
time delay changes at different observation epochs~\citep{Bonvin2018}.

In this paper, we ignore all these details and study on what level of the microlensing time delay variations would bias the results in LSST data.
In Section 2, we introduce the simulated LSST lenses and the lessons we learnt from TDC and H0LiCOW. In Section 3, we introduce the microlensing time delays.
The results are shown in Section 4 and we give the summaries and discussions in Section 5.
We assume a flat $\Lambda$CDM model with $\Omega_M=0.3$ and $H_0=70km/s/Mpc$ as the benchmark in the simulations.

\section{Lensing observation in LSST era}
The upcoming LSST will monitor $\sim 10^3$ strongly lensed quasars during its 10 years campaign repeatedly monitoring for 18000 $deg^2$ of the sky~\citep{OM10}.
To understand whether the proposed observation
strategy can provide sufficient information on time delay measurements, a Time Delay Challenge (TDC) program was conducted~\citep{Liao2015}. The challenge
``Evil" Team simulated thousands of time delay light curve pairs including all anticipated physical and experimental effects, while the
community was then invited to extract the time delays from the mock light curves blindly based on their own algorithms as the ``Good" Teams.
One of the main goals was to test the average precision and the bias to make sure the cosmological parameters can be constrained precisely and also accurately.
The TDC claimed that $\sim400$ well measured time delays would be obtained with precision $\sim3\%$ and bias $\leq1\%$. We call such lenses as ``Golden lenses".
We adopt this number of the ``Golden lenses" in our work as the benchmark, one can easily estimate the cosmological parameter uncertainties scaling as $1/\sqrt{N}$ if the number
of the ``Golden lenses" changes in the next challenge TDC2 that will make more realistic simulations.
On the other hand, to do cosmology we also need inputs from HST and Keck Telescopes
such that the quasar imaging and the central velocity dispersion of the lens can be got to do lens modelling.
However, in reality the observing time for them is a limited resource. It may take a long time for all LSST lenses applying in cosmology.

\begin{figure}
 \includegraphics[width=9cm,angle=0]{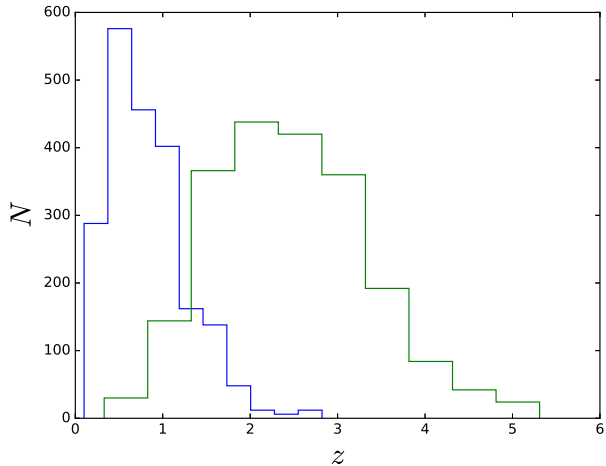}
  \caption{The redshift distributions of the lenses and sources from OM10 catalog where $\sim2000$ systems meet the detection criteria.
  }\label{z}
\end{figure}

\begin{figure}
 \includegraphics[width=9cm,angle=0]{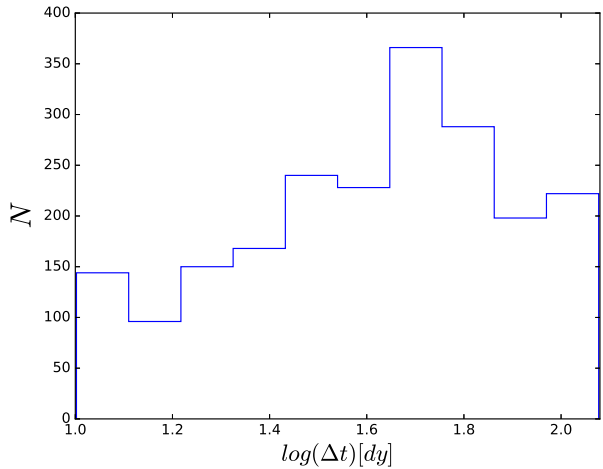}
  \caption{The time delay distribution, considering the quads as doubly imaged systems with the largest time delays.
  }\label{dt}
\end{figure}

In TDC, the selected time delays from the OM10
catalog~\citep{OM10} for simulation were between 10 days and 120 days and the image magnitudes were within the limiting magnitude 23.3 in the $i$ band, totally $\sim2000$ systems.
Though C18 claimed the
time delay ratios from quad systems can be well-constrained by the host imaging, providing the constraining
information on the microlensing time delays, there are more doubles in reality.
For LSST, only $15\%$ of the lenses are with quad images~\citep{OM10}. In principle, all three independent time delays in quads are used
to infer $H_0$ in the Bayesian framework. However, to make it simple, we only take the two images with the largest time delay which mainly dominate
the inference of the cosmology. Another consideration for this is based on the absolute microlensing time delay error which incline to bias
smaller time delays (see Sec.3), for example, the three close images in the cusp configuration. This assumption would result in 350 double-image-like time delays in this analysis.
The distributions of the redshifts and time delays from OM10 are plotted in Fig.\ref{z} and Fig.\ref{dt}. We notice that the lessons in TDC showed the precision of individual time delay measurements decreases with time delay itself consistent with
the time delay uncertainty being approximately constant in days~\citep{Liao2015}. This was expected that the absolute precision is mainly determined by the cadenced sampling
of the light curves. We take 1 day as the constant uncertainty so that the average precision is close to the TDC.

On the other hand, to infer $H_0$, we need the observation of high-resolution imaging of the lensed host galaxies, good PSFs to model the point images~\citep{Chen2016},
and the central velocity dispersion measurements to do the lens modelling and then determine the Fermat potential differences. According to the state-of-art
H0LiCOW program,
the lens modelling uncertainty for each lens can be achieved at percent level for individual systems with blind analysis controlling the systematics~\citep{Bonvin2017,Birrer2018},
we take the relative uncertainty $3\%$ as the benchmark in this work.

Finally, all mass along the LOS affects the lens potential that the light passes through. These mass fluctuations
can also lead to additional focusing and defocusing of the light rays, which in turn affects the measured time delays.
To avoid biasing the result,
one can give an estimate by spectroscopic/photometric observations of local galaxy groups and LOS structures in combination
with ray-tracing through N-body simulations, or even realistic simulations of lens fields~\citep{Collett2013,Greene2013,McCully2014,Treu2016}.
For HE0435-1223, the estimate of
external convergency $\kappa_{ext}$ would result in $2.5\%$ relative uncertainty on the time delay distance~\citep{Rusu2017}, we refer to this value
in our analysis.

\section{Microlensing time delays}
\begin{table*}\centering
 \begin{tabular}{lccc}
  \hline\hline
  $\sigma_{\Delta\phi^{Lens}}/\Delta\phi$ & $\sigma_{\Delta\phi^{LOS}}/\Delta\phi$ & $\sigma_{\Delta t^{LC}}$ & $\sigma_{\Delta t^{ML}}$ \\
  \hline
  $3\%$  & $2.5\%$  & $1\ day$ & $1/3,\ 1,\ 3\ days$  \\
  \hline\hline
 \end{tabular}
 \caption{Uncertainties of the related observations for time delay cosmology.
}\label{error}
\end{table*}

The community used to believe that the time delays measured from the light curve pairs
only depend on the large structure of the lenses and can be directly used in constraining cosmological parameters.
The systematics only come from the independent microlensing magnification light curves for each image and the algorithms to extract the time delays. However, TK18
questioned the use of measured time delays in cosmology. They proposed that an important physical process had not been
considered, that is, due to the finite disk size and the differential magnification of the coherent temperature fluctuations,
the microlensing effect changes the time delays on the scale of the light crossing time of the accretion disc ($\sim$days). While
the accretion disc moves relative to the lens, the microlensing time delay also changes.

Therefore, to get an unbiased result, one
needs to understand the details of all physical processes, which is hardly achieved currently. Although C18 and B18 have considered
the microlensing time delay effects in a Bayesian framework, there still exist a lot of uncertain inputs for the microlensing time delay priors.
First, the assumed ``thin disc" and ``lamp-post" model~\citep{Cackett2007} on accretion disc may not be correct. The size of the accretion disc may be larger
than the prediction from the standard thin disk theory with Eddington ratio 0.1~\citep{Shappee2014,Mosquera2011,Morgan2010,Kollmeier2006,Fausnaugh2016}, which would make the systematics severer. Alternative accretion disk models giving different variability are proposed~\citep{Dexter2011}.
Second, the adopted fixed best-fit local environments for images and mass function for the stars
can also bring uncertainties~\citep{Chen2018}. For different local convergency $\kappa$, shear $\gamma$ and star proportion $f_*$, the standard deviation of the magnification
map changes~\citep{Liao2015}, so does the microlensing time delay map. Third, the inclination, position angle and especially the size of the accretion disk make the time delay map
different~\citep{Tie2018}, while C18 has considered a conservative range of these. At last, the microlensing time delay changes with the relative motion of the source, while C18 and B18 considered
the time delays from three finite epochs as constants. Whether it is appropriate for LSST 10-year light curves needs further test.

Nevertheless, from the analysis of RXJ 1131-1231, HE 0435-1223 and PG 1115+080~\citep{Tie2018,Bonvin2018,Chen2018}, the microlensing time delay for individual images
is estimated to be from $\sim0.1$ to $\sim4$ days, so is the one between two images, and it is an absolute rather than fractional error. Therefore,
to assess the impact of the microlensing on LSST time delay cosmology, we assume 1/3, 1 and 3 days, respectively
as the microlensing time delay uncertainty between two images for all LSST lenses.

\section{Methodology and results}

According to the strong lensing theory, the cosmological time delay between two images $i, j$ is given by:
\begin{equation}
\Delta t^{COSM} = \frac{D_{\mathrm{\Delta t}}(1+z_{\mathrm{d}})}{c}\Delta \phi, \label{timedelay}
\end{equation}
where $c$ is the light speed.  $\Delta\phi=[(\boldsymbol{\theta}_i-\boldsymbol{\beta})^2/2-\psi(\boldsymbol{\theta}_i)-(\boldsymbol{\theta}_j-\boldsymbol{\beta})^2/2+\psi(\boldsymbol{\theta}_j)]$
is the difference between Fermat potentials (representative of $\langle\kappa\rangle$ mentioned above) at different image angular positions $\boldsymbol{\theta}_i,\boldsymbol{\theta}_j$, with $\boldsymbol{\beta}$ denoting the source position, and $\psi$ being  the two-dimensional lensing potential determined by the Poisson equation $\nabla^2\psi=2\kappa$, where $\kappa$ is the surface mass density of the lens in units of the critical density $\Sigma_{\mathrm{crit}}=c^2D_{\mathrm{s}}/(4\pi GD_{\mathrm{d}}D_{\mathrm{ds}})$,   $D_{\mathrm{d}}$, $D_{\mathrm{s}}$ and $D_{\mathrm{ds}}$ are  angular diameter distances to the lens (deflector) located at redshift $z_{\mathrm{d}}$, to the source located at redshift $z_{\mathrm{s}}$ and between them, respectively.

From the observation aspect, the determination of Fermat potential difference can be split into two independent parts:
\begin{equation}
\Delta \phi=\Delta \phi^{Lens}+\Delta \phi^{LOS},
\end{equation}
where $\Delta \phi^{Lens}$ is determined by lens galaxy observation and $\Delta \phi^{LOS}$ is dermined by mass distribution along the line of sight.
People used to only consider the lens structure. When we give the measurements of the LOS environment, we can take it as the statistical
uncertainty rather than the systematics. As the same, the time delay
also include two parts:
\begin{equation}
\Delta t^{COSM}=\Delta t^{LC}-\Delta t^{ML},
\end{equation}
where $\Delta t^{LC}$ is from the light curves and $\Delta t^{ML}$ now corresponds to the microlensing time delay. If we can understand the detailed
physical processes and have high-quality observation to assess $\Delta t^{ML}$, we can take it as the statistical one as C18 and B18 suggested. Otherwise, this would be
an important systematic uncertainty source.

We consider three cases that may happen for the community in this work.
Case 1: considering the AGN model has not been well understood, there is a scenario that the microlensing
effect is small enough which could be ignored. We study $H_0$
constraint based on simulated LSST data. Case 2: we consider the condition that the microlensing time delays can be well estimated as the statistical errors or priors in Bayesian framework,
we also give the constraint results. Case 3: the effects do exist but one may neglect it when he does the analysis, this would underestimate the uncertainty and
result in an biased $H_0$ estimation. We test how reliable this could be by studying the induced systematics on $H_0$.

To avoid the impact by specific noise realisation, we adopt the minimum $\chi^2$ statistics in the analysis using different noise realisations for the observables.
The statistic is expressed as:
\begin{equation}
X^2=\sum_{i=1}^{N}\frac{[ D_{\Delta t,i}^{obs}-D_{\Delta t}^{th}(H_0,\Omega_M;z_{d,i},z_{s,i})]^2}{\sigma_{D_{\Delta t,i}}^2},
\end{equation}
where
\begin{equation}
\sigma_{D_{\Delta t,i}}^2=\frac{c}{1+z_{d,i}}(\frac{\Delta t_i^2\sigma_{\phi,i}^2}{\Delta\phi_i^4}+\frac{\sigma_{\Delta t,i}^2}{\Delta\phi_i^2})+(D_{\Delta t,i}^{obs}*2.5\%)^2.\label{g}
\end{equation}
Note that we use $X^2$ rather than the conventional $\chi^2$ to reminder the readers the uncertainty given by Eq.\ref{g} is not
a rigorous Gaussian distribution since it is through error propagation from Tab.\ref{error}. However, we still adopt Guassion
assumption in the analysis. First, rather than inferring an accurate $H_0$ from realistic data, we mainly give an estimate on
the impact of microlensing time delays, the relative impact level is what we are interested. Second, to correctly assess the bias
by non-Guassian effect, one has to start from the original observations, i.e., the pixel values of the host imaging, the velocity dispersion,
the AGN positions, and the time delays taken as Guassians. However, these observational uncertainty details have not been set up for LSST lenses.
Third, the H0LiCOW shows the inferred $D_{\Delta t}$ and $H_0$ approximately follow Guassian distributions~\citep{Wong2017}.
Therefore, while the analysis can be presented simply and clearly for the readers, we think the main conclusion would not change.
This assumption also appears widely in the literature~\citep{Paraficz2009,Coe2009,Linder2011}.
More details need further studying in the ongoing TDC2 (being prepared by TDC team) and TDLMC programs~\citep{Ding2018}.

\begin{figure}
 \includegraphics[width=9cm,angle=0]{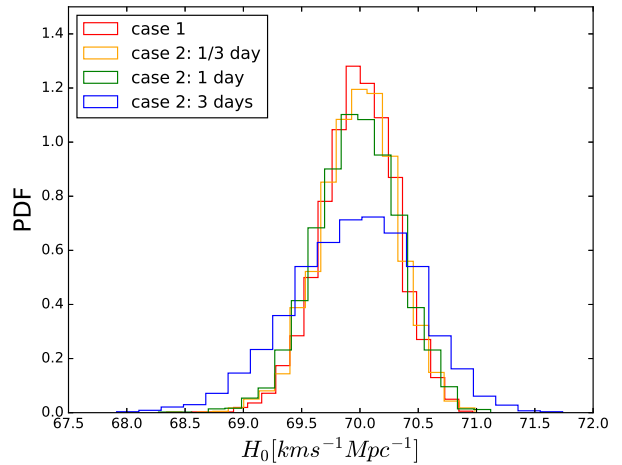}
  \caption{Given specific 350 systems, 1-D marginalized PDF of $H_0$ for different microlensing time delay uncertainties as statistical inputs.
  }\label{stats}
\end{figure}

\begin{figure}
 \includegraphics[width=9cm,angle=0]{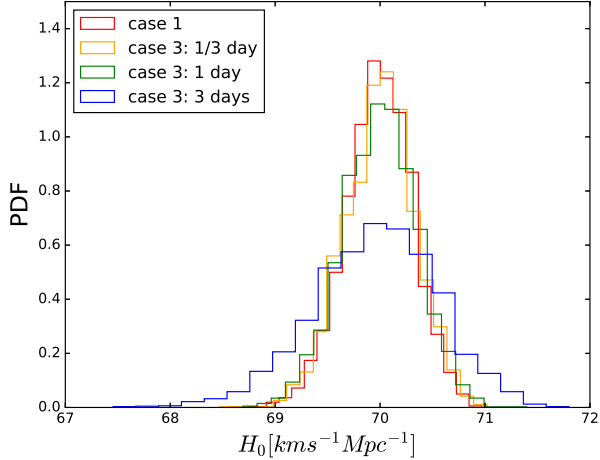}
  \caption{Given specific 350 systems, 1-D marginalized PDF of $H_0$ for different microlensing time delay uncertainties as systematical inputs. The PDFs include
  both statistical and systematical uncertainties.
  }\label{sys}
\end{figure}

Given the randomly selected 350 lensing systems, we distribute noises to Fermat potential
differences, LOS masses, time delays from the light curves and microlensing time delays as summarized on Tab.\ref{error}. For each noise realisation, we do minimization
to find the best fit values of $H_0$ and $\Omega_M$. We repeat this process for 3000 times for different noise realisations and take all best-fit values
as the constraint inputs, the marginalized histograms of $H_0$ are plotted in the Fig.\ref{stats} and Fig.\ref{sys}.
They approximately look like Gaussian distributions. Thus,
we simply calculate the standard deviations of the PDFs as $1\sigma$ uncertainty.
Furthermore, to avoid the impact of specifically selected systems, we repeatedly select 350 systems from whole OM10 catalog consisting of $\sim2000$ systems that meet
LSST criteria~\citep{Liao2015} for 15 times and for each dataset, we repeat the above process.
Finally, we calculate the mean of all uncertainties as the average constraint power on $H_0$ summarized in Tab.\ref{result}.
For case 3, the standard deviations must be larger than case 1, which
can be seen as the combinations of statistical and systematical uncertainties . It is due to the microlensing time delays as the systematics that enlarge the variations.
We subtract the statistical ones in case 1 and get the systematics/bias of $H_0$, see case 3-1 in Tab.\ref{result}. As we can see from the results, the microlensing time delay matters when it
is typically larger than 1 day, when the systematics is comparable to the statistical uncertainty, leading to a biased estimate on $H_0$.

\begin{table*}\centering
 \begin{tabular}{lcccc}
  \hline\hline
   & & 0.3 day & 1 day & 3 days \\
  \hline
  case 1 & $0.45\%$ & & &\\
  case 2  &  & $0.47\%$ & $0.51\%$ & $0.76\%$ \\
  case 3-1  &  & $0.12\%$ & $0.22\%$ & $0.70\%$  \\
  \hline\hline
 \end{tabular}
 \caption{Relative uncertainties $\sigma_H/H_0$ for case 1, 2 and case 3-1, i.e., subtracting
 the statistical one in case 1 from case 3.
}\label{result}
\end{table*}

\section{Conclusions and discussions}
We have tested the robustness of future LSST strong lens time delay cosmology by studying the systematics in the time delay measurements.
We summarize the main conclusions as follows:

1) With the assumptions that all current measurements related with strong lens time delay cosmology are accurate including
the time delays from the light curves, i.e., the scenario where microlensing time delays are ignorable,
$\sigma_H/H_0$ can be constrained to 1$\sigma$ uncertainty $0.45\%$ for 350 lenses in LSST.

2) Cosmological time delay measurements may be affected by microlensing effects. If these microlensing time delays can be estimated correctly as extra
statistical uncertainties or priors in the Bayesian framework, the constraint on $H_0$ will be weaker as expected.
We assume 1/3, 1 and 3 days respectively as the typical microlensing time delay uncertainties for all
lenses, the 1$\sigma$ uncertainties of $\sigma_H/H_0$ are $0.47\%$, $0.51\%$ and $0.76\%$, respectively.

3) If the microlensing time delays exist but we ignore them, they could be the systematics, i.e., one would underestimate the observational uncertainties in inferring
$H_0$. Taking $0.45\%$ as the expected average statistical uncertainty for $\sigma_H/H_0$, we show the systematics of $0.12\%$, $0.22\%$, $0.70\%$, respectively inside the
$H_0$ estimation for 1/3, 1 and 3 days of the microlensing time delay uncertainties. Therefore, microlensing time delay larger than 1 day would strongly bias
the result. If the systematics is small enough, the lens
modelling and LOS uncertainties would dominate the total uncertainty.
For current tension $\sim7\%$ between Strong Lensing $H_0\sim72km/s/Mpc$~\citep{Birrer2018} and CMB $H_0\sim67km/s/Mpc$~\citep{Planck2016},
we propose a hypothesis that it may partially stem from the microlensing time delay errors (depending on how large they are).
For example, for a $\sim$14 days time delay like
in HE 0435-1123 or PG1115+080, 1 day microlensing bias would result in a tension at this level, though for the very 4 quad systems~\citep{Birrer2018},
it is not likely to result in a overall $2\sigma$ tension.

For future study, we need more knowledge on AGN accretion model from astrophysics inputs and more precise measurements for local image environments.
If we can confirm that the microlensing time delays are typically smaller than 1 day or much smaller than statistical uncertainties like in Birrer et al. 2018, the result would be seen as unaffected.
The relative motion of the source and the monitoring time also matter. If the images stay locally on the microlensing time delay map, the
microlensing time delays can be considered as constants. This can strongly bias the results especially for the point lying close to a micro caustic.
On the other hand,
if the image range is large on the map, the mean microlensing time delay may be close to zero but still non-ignorable,
we could use either a constant for the whole 10-year light curves or
epoch-dependent model for describing it, which needs
further discussions.
Selecting the lenses with larger time delays like SDSS 1206+4332~\citep{Birrer2018} may make the absolute systematics less important, for example, we have made an estimate,
there are 1/4 of the systems with time delays larger than
60 days, the statistical uncertainty of $H_0$ is $0.81\%$ and the systematics are almost the same.

We may also extend the strong lensing systems by using other sources, for example, the supernovae are point sources such that the time delay measured are cosmological~\citep{Kelly2015}.
Furthermore, the transients like gravitational waves~\citep{Liao2017b} can also measure the time delays precisely and accurately.

At last, note that we only test the systematics of time delays in this work, in addition, the lens modelling and LOS should also be tested to understand the systematics
therein. The results in this work should not be simply seen as the prediction of LSST cosmology. The systematic floor needs to
be further studied to let us know how powerful on earth the lensing method could be in future cosmological studies.

\section*{Acknowledgments}
The author thanks the referee for his/her valuable suggestions and Xuheng Ding for polishing the paper.
This work was supported by the National Natural Science Foundation of China (NSFC) No. 11603015
and the Fundamental Research Funds for the Central Universities (WUT:2018IB012).

\clearpage

\end{document}